\documentclass[prd,aps,preprint,tightenlines,superscriptaddress,showpacs]{revtex4}
\usepackage{epsfig}
\usepackage{amsmath}

\newcommand{\beq}[1]{\begin{equation} \label{#1}}
\newcommand{\eeq}{\end{equation}}
\newcommand{\beqar}[1]{\begin{eqnarray} \label{#1}}
\newcommand{\eeqar}{\end{eqnarray}}
\newcommand{\beqs}{\begin{eqnarray}}
\newcommand{\eeqs}{\end{eqnarray}}

\newcommand{\nid}{\noindent}

\newcommand{\bib}{\bibitem}

\newcommand{\vtab}{\vspace*{0.5cm}}

\begin{document}

\title{{\Large Braving the Challenges of Being Small}
\footnote{Talk given at the Education Forum of OCPA04, Joint
International Conference of Chinese Physicists Worldwide,
Shanghai, PRC.}
\\

\vtab

--The Unique Physics Program at Manhattanville College}

\vtab

\author{Zhang Chen}
\email{chenz@mville.edu}
\affiliation{Department of Physics, Manhattanville College, Purchase, NY 10577, USA}

\date{\today}

\begin{abstract}
\nid We present the formulation and development of the unique
physics program at Manhattanville College. By addressing
challenges faced by small physics departments in liberal arts
colleges, such as the lack of critical mass, lack of significant
funding, and under-staffing, we share our efforts and experiences
in revitalization of the program into one that is small but of
rigor and vitality, with unique features that take advantage of
the resources available, including adoption of computer technology
in instructional laboratory, collaborations with nearby
institutions, and joint efforts with colleagues of other
departments/schools. Also included are some discussions on the
balance between service and the major, on the relationship with
K-12 physics education, and on the rigor of the curriculum,
particularly in view of the opposite but converging trends in US
and in China.
\end{abstract}

\pacs{01.40.-d, 01.50.Lc}


\maketitle

\section{Introduction}

The outline of the paper is as follows. In Section
\ref{sec:Background}, we give a brief introduction to Manhattanville
College and its background. In Section \ref{sec:Program} we provide
a general overview of the physics program/department at Manhattanville.
Section \ref{sec:Unique} contains a more detailed discussion of
several special features of the program that make it unique.
Section \ref{sec:Future} presents some ideas and plans of future
growth as well as some more general discussions on the trends
in physics higher education in US and China. We conclude the paper
by giving the summary and outlook in Section \ref{sec:Summary}.

\section{Background and Current Status of Manhattanville} \label{sec:Background}

Manhattanville College was founded in 1841, originally as a Catholic
all-female college. In Spring 1971, it became fully co-educational
and non-denominational. Currently it is in essence a residential
private liberal arts college.

Manhattanville was initially located in the Manhattanville area
of Manhattan, New York City. It later moved into Purchase, Westchester
County, NY, about $25$ miles north of NYC, a very convenient and
attractive location.

The College now has ~$1,500$ full-time undergraduate students,
of which about 85\% are residential. The student body
represents more than three dozens of states and fifty nations.
The acceptance rate is at about fifty per cent, with average
incoming SAT ~$1100$ and GPA ~$3.0$. More information can be found
at the college's web-site at http://www.mville.edu/


Manhattanville College's mission is to "educate socially and
ethically responsible global leaders". Its general education
vision is to provide every student a broad-based liberal arts
education, with highly attentive and individualized instructions.
As is indicated in New York State regulations, a student need
$120$ credits to graduate, and $90$ of those credits have to
be from liberal arts courses for a Bachelor of Arts (BA) degree.
A student can get a Bachelor of Science (BS) degree with $60$
liberal arts credits (among the total $120$)\cite{ref:NYS}.

At Manhattanville, in addition to complying with New York State
regulations, we have a set of requirements to further
assure that our students come out with well-rounded skill sets:
A student needs both a major AND a minor to graduate;
he/she needs to fulfill a distribution requirement that he/she
must take at least $6$ credits each in four area out of the
following five: Science \& Mathematics, Social Sciences,
Humanities, Arts, and Foreign Languages; a student must also
fulfill a Global Awareness requirement to broaden his/her
visions and understandings beyond his/her own original
cultural background.

All these requirement are set up in a well-developed framework
called the "Portfolio System", in which the students set out
to plan his/her college education together with their advisors
from the beginning and have constant checks and adjustments on
their plans through out the course of study. This process starts
with the freshmen "Preceptorial", a one-year general liberal arts
course in essence being "introduction to college learning
and life", and then continues till graduation. This structured
and flexible approach greatly helps the students along their
higher education journey.

\section{Physics Program at Manhattanville: An Overview} \label{sec:Program}

Manhattanville has had a physics program for a very long time.
Before the 70's, it was a full undergraduate program with a set
of courses offered and both major and minor students. It had a
stable faculty size and would graduate several majors each year.
The instructional laboratory was also decently established.

However, since the change of nature of the College in the 70's,
the physics program went into a continuous declination, till
the summer of 1999. At the time, there was only one course offered
each semester (the pre-med algebra-based introductory course),
the program had no major or minor student(s), the department
had zero full-time faculty (the one course was taught by
a part time adjunct faculty member), and there had been no
new purchase or update of any laboratory equipment.

Starting fall of 1999, with the hiring of the first full-time
faculty member in physics in many years, the physics program
at Manhattanville underwent a revitalization process. We
completely redesigned the whole curriculum, purchased and
updated the equipment in the instructional laboratory in a
continuous fashion, devoted much efforts in raising the
awareness of the program both on and off campus, and attempted
different innovative ways in making the program more attractive
by tapping into its strengths and circumvent its weaknesses.

As of spring of 2004, the physics program at Manhattanville
has become a small but efficient one, with offerings of
three different introductory physics courses simultaneously
and a set of intermediate physics courses alternating between
semesters and academic years. With a stable physics minor student
body and the emerging major enrollment, advanced level courses
will also soon become a regular part of the course offerings.
On the facility side, we have updated and rebuilt much of the
introductory instructional laboratory with newly purchased
instruments. We have also developed a computer-based instructional
laboratory and a summer research internship, both to be covered
in more detail later, to provide the experimental training
students need. Currently the program is running smoothly and
successfully at full capacity with one full-time and one
part-time adjunct faculty members. It is posed in a ready
position to develop further once the situations in funding
and staffing are ready.

\section{Special Features of the Physics Program}\label{sec:Unique}

The Physics Program at Manhattanville has many unique features that
set it apart from any other program in the country. First of course
is its small size \cite{ref:chairs}, which is both the source of its
strength and uniqueness and the origin of the challenges and
difficulties it faces. We tried to develop a program that utilizes any
and all the resources that we could draw upon, and have indeed
been successful in achieving a nice balance under the many constraints.

\subsection{The Major Program}

The redesigned physics major requires $54$ mandatory credits total,
which includes a large $16$-credit mathematics co-requisite. This
means the physics requirement is $38$ credits, quite in line with
other liberal arts (including science) majors at the college. It
is expected that the requirements could be finished reasonably by
a student in either four or (even) just three years.

The required course structure is: one year of calculus based
introductory physics sequence (University Physics I \& II,
$ 2 \times 4 \, cr $); the intermediate physics core consisting
of four $2000$ level courses (Mechanics, Electromagnetism,
Quantum Physics, and Thermodynamics \& Statistical Physics,
$ 4 \times 4 \, cr $); two advanced electives chosen from
a set of $3000$ level special topic courses (eg, Advanced
Mechanics, Quantum Mechanics, Solid State Physics, etc,
$ 2 \times 4 \, cr $); one summer research internship
( $ 3 \, cr$); and the senior research seminar with a thesis
( $ 1 \times 3 \, cr$ or $ 2 \times 1.5 \, cr$). The mathematics
courses are Calculus I, II \& III, and Ordinary
Differential Equations ($ 4 \times 4 \, cr$). There is also
a minor program that requires the one year introductory
sequence, two intermediate physics courses, and a one
year calculus sequence (Calc I \& II). Students are encouraged
to take more advanced mathematical courses like Linear Algebra
or Complex Variables, some physical or other science courses
like General Chemistry and/or Biology to deepen and/or broaden
their scientific background.

With a program set up like the above, the students, while
receiving rigorous training, also have the freedom to adjust
and streamline their course of study to match their own general
study plan (the "Portfolio", see \ref{sec:Background}).
They can, for example, choose the timing of the summer internship,
even opt to do an extra one; pace themselves a bit by taking
just one intermediate core course a semester while fulfilling other
college requirements or pursuing their other interests in liberal
arts; or structure their senior research in either an intense
one semester or a thorough two semester fashion. Having both
high-standard courses and different structural options, the
major/minor program at Manhattanville is one of rigor and flexibility.

\subsection{The Service Courses}

Besides the rigorous calculus based introductory physics sequence,
the department also offers on a regular basis two other one-year
introductory physics courses. One is the also rigorous algebra
based "College Physics" sequence, mainly intended for students
in various pre-health programs and/or biological sciences. It
in fact has the largest enrollment of all physics courses at
Manhattanville, and is the most important service course of the
department.

The other service course is a conceptual introduction to physics
and astronomy intended for general liberal arts students. Such a
course engages students from the broadest spectrum and focuses on
three aspects:

\begin{itemize}

\item{The methodology of physics: Illustrating the scientific
approach to a vast array of different problems.}

\item{The knowledge body: Going over major developments of mechanics
and modern astronomy. Emphasis is on basic ideas and concepts rather
than techniques of problem-solving.}

\item{Relationship to everyday-life: Giving the students an
appreciation of how to evaluate the merit (or lack of merit) in
arguments presented in current scientific controversies, such
as debates about the environment, about energy usage and conservation,
about evolution, etc.}

\end{itemize}

\nid The sequence has been a success since its inaugural year
in 2000-2001.  It is now being offered each year and is taught
by a part-time adjunct faculty.

Indeed, all three introductory sequence at Manhattanville have stable
enrollment. We are pleased to be able to offer them to meet
the needs of our diverse student body within such a small department.

\subsection{The Summer Internship}

The summer research internship mentioned above is a unique feature
of the physics program at Manhattanville. Due to the status of the
instructional laboratory, staffing and budgetary constraints at the
college and departmental levels, full fledged intermediate and/or
advanced laboratory courses as part of the curriculum are unpractical
at least for the moment. On the other hand, the convenient location
of Manhattanville--close proximities to New York City and major research
institutions therein--brings about many opportunities.

We took advantage of the close ties between our faculty members and
those at Physics Department of Columbia University in NYC and established
a collaboration with Columbia. Our major students will take summer
research internship course(s) in the summer(s) after their sophomore year
and/or junior year as supplements to the theoretical courses. The
students will be engaging in actual research activities
at the Nevis Research Laboratory of Columbia University, located in
Irvington NY, even closer to Manhattanville, by joining as interns
one of the groups working there.

Such experiences will be extremely beneficial to the development of
students in many ways. They are being put into an active scientific
environment that facilitates interactions with researchers and peer
students. They are also being exposed to modern experimental instrumentation,
techniques and methods.  The summer research internship will be an important
and effective method for the students to obtain the necessary
laboratory experiences and training, in lieu of the conventional
intermediate and advanced laboratory courses. In the long run, it
will be an unique and integral part of the whole experimental training
aspect of the program.

\subsection{The Computer-based Laboratory}

One of the innovative ways in physics education has been the
application of computer technology in course instruction. We
have started to develop a computer-based instructional laboratory
component in addition to conventional methods. Students will be
engaging in computer-based data acquisition and analysis and performing
computer simulated intermediate and advanced experiments otherwise
not accessible to them. This is another effective and innovative
method of providing the needed training under our constraints.

Currently we have three laboratory stations set up, each with a
PASCO physics (USB) 750 interface bundle and a Dell small-form
desktop computer. They are going to be fully integrated into
the curriculum starting fall of $2004$, with students performing
experiments using them in rotation. This puts us in a leading
position among comparable institutions on this aspect, and in
the long term will be another integral and complementary part
of the physics program.

\section{Opportunities and Challenges} \label{sec:Future}

The Physics program is currently running at its full
capacity efficiently. Besides the continued efforts in trying
to increase enrollments of major and non-major students
and in upgrading, updating and expanding the laboratory,
we are also looking ahead for opportunities of growth and
further development.

One such great opportunity is collaboration with our School
of Education. Higher standards from New York State Department
of Education for future teachers (K-12) are being implemented
in schools throughout the state, representing a national trend
of raised awareness of the importance of better teacher education,
especially in sciences. The new standard require disciplinary
content equivalent to a undergraduate major, and in particular
for physics, $32$ credits (including $8$ in mathematics).
The School of Education at Manhattanville is very successful
with great reputation and track record. We have taken advantage
of this opportunity and developed, in collaboration with the
School of Education, both an undergraduate educational track
physics major and an Mater of Arts in Teaching (MAT) program in
physics. These may lead to potential significant numbers of
students for the physics program in the upcoming years.

Another mid-long term planning of our program has development
of interdisciplinary programs in mind. The 21st century has
seen the emergences and rapid expansions of many interdisciplinary
fields, as well as the ever-growing significance in fields
like biological and material sciences. We want to tap into
the strengths of existing programs at Manhattanville and
develop new and exciting interdisciplinary programs. One such
example would be a program in applied mathematical physics.
Manhattanville has excellent programs in both computer
science and mathematics. By collaboration with the Department
of Computer Science and Mathematics, we plan to develop a strong
and appealing program that provide students with knowledge
in physics and mathematics, abilities in abstract and
quantitative analysis, and skills in applied mathematics
and computer science. Such a combination will be
an attractive option for the students, facing the
technology intensive economy and job market. Other
programs being considered including bio-physics,
bio-informatics, and chemical physics.

The biggest challenges facing the Department are two-fold.
First, the lack of critical mass in both the faculty and
the students. We are actively engaged in various methods
of recruiting and retaining students, hoping to further
increase our student body. At the same time, the eventual
increase of faculty size is also a priority of the
department. The second is lack of significant development
funding. While we have achieved a lot under current
budget constraints, major growth in facilities and/or
faculty size still need major funding. We expect to receive
continued support from the college administration on these
efforts, at the same time of actively seeking external funding
as well.

\section{Summary and Outlook} \label{sec:Summary}

The Physics Program at  Manhattanville is one
of unique features and strengths--the collaborative summer
research internship and the computer-based instructional
laboratory being two, despite its small size.
Students in the program will receive rigorous and highly
individualized training in both theoretical and experimental
aspects of the discipline.

On a bigger scale, enrollment in physics courses in K-12
education has started to increase\cite{ref:chair04}. This
is an indication of a national trend in the heightened
importance people starting to assign to sciences and
physics in particular. The direct result is the potential
larger enrollment for physics programs and the requirement
of more rigorous curricula in physics. On an even larger scale,
globally sciences and physics in particular are, more than ever,
recognized as the underlying driving force of development
and advancement in economics and technology. The theme
is to find the best way of educating students in seeking
the balance of knowledge body and methodology. China, for
example, is starting to turn away from its traditional
high-intensity, large-depth, narrow-focus approach of
physics undergraduate education, and beginning to experiment
in a more broad-based, liberal arts oriented approach.
For example, the "Science Platform" at Fudan University\cite{ref:chn},
one of the top comprehensive universities in China, is
such an attempt by requiring all science majors to take
biology, physics and chemistry in their first year.
It is not clear yet what the best balance is, but this
presents opportunities for discussion, research and
collaboration within the physics community, or even
the whole science community, at a global scale.

We are confident that with the increasing awareness of
the physics program, the continued pursuit of excellence
and innovation, and the constant seeking of collaboration
and cooperation, the program will, riding the favorable
trend statewide, nationally and internationally, expand
and grow and have a bright future.

\vspace*{1cm}

\nid \acknowledgements

\nid The author was partly supported by a research grant from
Manhattanville College. The author would like to thank the
hospitality of Prof. Ruibao Tao and the Physics Department
of Fudan University, Shanghai, PRC. The author would also like
to thank Laura Wang, without whose encouragement and support
this work would not have been possible.




\end{document}